\documentclass[preprint]{revtex4}
\usepackage{graphicx}
\usepackage{color}

\begin{document}
\title{ From companies to colonies: The origin of Pareto-like distributions in ecosystems  }
\author{Alon Manor and Nadav M.  Shnerb }
\affiliation{ Department of Physics, Bar-Ilan University, Ramat-Gan
52900 Israel  }

\begin{abstract}
Recent studies  of  cluster distribution in various ecosystems
revealed Pareto statistics for the size of spatial colonies. These
results were supported by cellular automata simulations that yield
robust criticality for endogenous pattern formation based on
positive feedback. We show that this self-organized criticality is a
manifestation of  the law of proportionate  effect, first discovered
in the context of business firm size. Mapping the stochastic model
to a Markov birth-death process, the transition rates are shown to
scale linearly with cluster size. This mapping provides a connection
between   patch statistics and the dynamics of the ecosystem; the
"first passage time" for different colonies emerges as a powerful
tool that discriminates between endogenous and exogenous clustering
mechanisms.  Imminent catastrophic shifts (like desertification)
manifest themselves in a drastic change of the stability properties
of spatial colonies, as the chance of a cluster to disappear depends
logarithmically, rather than linearly, on its size.
\end{abstract}

\maketitle

\def\baselinestretch{1}

In spatially extended ecosystems the observed patchiness supplies a
lot of information on a given  system's parameters and may signal
that the system is approaching a transition \cite{rit}. Of
particular interest  are ordered \cite{o1,o2,o3,o4} and disordered
\cite{oz} vegetation patterns in the semi-arid climatic zone, where
climate changes and overgrazing lead to an accelerated process of
desertification \cite{global}. Recently, two groups of researchers
simultaneously published  new results about patch-size statistics.
Using high-resolution remote sensing techniques, these groups found
power-law distributions of vegetation patch size (the probability of
finding a vegetation patch of size $n$ falls like $n^{-\beta}$ where
$\beta$ is greater than unity and can vary between different
regions) along the Kalahary Transect \cite{scanlon}, and in three
Mediterranean ecosystems \cite{kefi}. Another new study
\cite{vandermeer} has shown that the spatial distribution of ant
colonies in a tropical agroecosystem is also Pareto-like. Power-law
statistics indicate the absence of an intrinsic scale in the system,
and in many cases its appearance signals that a stochastic system is
"at criticality", i.e., experiencing a continuous phase transition
\cite{solbook}. It is surprising to find scale-free distributions
along a wide range of environmental conditions; in most systems it
appears only at the transition point \cite{sole}. The challenging
problem, thus, is to identify the mechanism that yields  robust
critical behavior for different ecosystems under a wide scale of
environmental conditions.

To support and explain their findings, all three groups suggested
(slightly different) cellular automata models that support self
organized criticality. Effectively, there are  only two parameters
in these models:  the global constraint on the overall fractional
tree cover, $f^*$, and the local strength  of positive-feedback
between plants (for the sake of concreteness we use hereon the
spatial vegetation terminology; our results, however, are generic).
Facilitation implies that the chance of an occupied cell to become
empty decreases, and the chance of an empty cell to become occupied
increases, if its neighboring cells are occupied. It turns out that
these two constituents are sufficient to ensure self-organization
and power-law distribution of patch sizes, while the details of the
model (like the definition of neighborhood, kernel function, etc.)
affect only the slope  $\beta$. These results increase the need to
identify the underlying feature that makes criticality  a robust
property of models with global capacity constraints and
positive-feedback.

In order to elucidate the dynamical law that governs self-organized
criticality, we have mapped the cellular automata used by Scanlon
et. al. into a Markov chain. The cellular automaton introduced  by
\cite{scanlon} was used here with only slight modifications. Every
cell is either empty (0) or covered by tree canopy (1), and local
facilitation determines the transition probabilities between these
two states. The  actual overall fraction of occupied cells, $f$,
affects the dynamics if it deviates from the fractional tree cover
allowed by the environmental conditions (the carrying capacity)
$f^*$. The transition rates in the $(i,j)$ cell are,
\begin{eqnarray}
  P_{i,j}(0 \rightarrow 1)&=&\rho_{i,j} + \frac{f^* - f}{1-f} \nonumber \\
  P_{i,j}(1 \rightarrow 0)&=&\rho_{i,j} + \frac{f - f^*}{f},
\end{eqnarray}
where $\rho_{i,j}$ reveals the effect of facilitation by other
biomass units in the local neighborhood,
\begin{eqnarray}\label{rho}
    \rho_{i,j}= \frac{1}{S} \sum_{k,l \in S} \sigma_{k,l},   \quad
    \sigma_{k,l}=\left \{
    \begin{array}{ll}
    1 & \mbox{if site ($k,l$) is populated}\\
    0 & \mbox{if site ($k,l$) is empty}
    \end{array}
 \right.
\end{eqnarray}
where $S$ is the area of a region of diameter $r$ centered at the
$(i,j)$ cell. Grazing was added to the model  by letting a fraction
$G$ of the "death" events  occur at a constant rate,  independent of
the local density $\rho_{i,j}$.

During the simulation, clusters were tracked and their sizes ware
monitored. The birth and death rates for size $n$  clusters were
determined by dividing the number of transitions $n\rightarrow n+1$
and $n\rightarrow n-1$, respectively, by the average number of size
$n$ patches. The parameters $\gamma$,$\  \alpha$ and $\Delta_0$ were
extracted from  $b_n$ and $d_n$ using linear regression.

 Switching to  patch size oriented
dynamics, the events where a patch of size $n$ is grew or decreased
were monitored to yield the corresponding transition rates. It turns
out that a stochastic birth-death process (where only $\pm 1$
transitions are allowed) is a remarkably good approximation for the
real dynamics. If $b_n$ is the birth rate (the chance of a patch of
$n$ cells to grow by one cell, or one vegetation unit, to $n+1$) and
$d_n$ is the corresponding death rate, the time evolution of the
patch size distribution $P_n(t)$ is described by the master
equation:
\begin{equation}\label{rateEquation}
\frac{d}{dt}
P_n(t)=-(b_n+d_n)P_n(t)+d_{n+1}P_{n+1}(t)+b_{n-1}P_{n-1}(t),
\end{equation}
and its steady state $\pi_n$ should satisfy the local balance
equation $\pi_n b_n = \pi_{n+1} d_{n+1}$. This allows for a
recursive solution for the steady state  \cite{stoch}, once the
birth-death transition probabilities are given.

\begin{figure*}
\includegraphics[width=13cm]{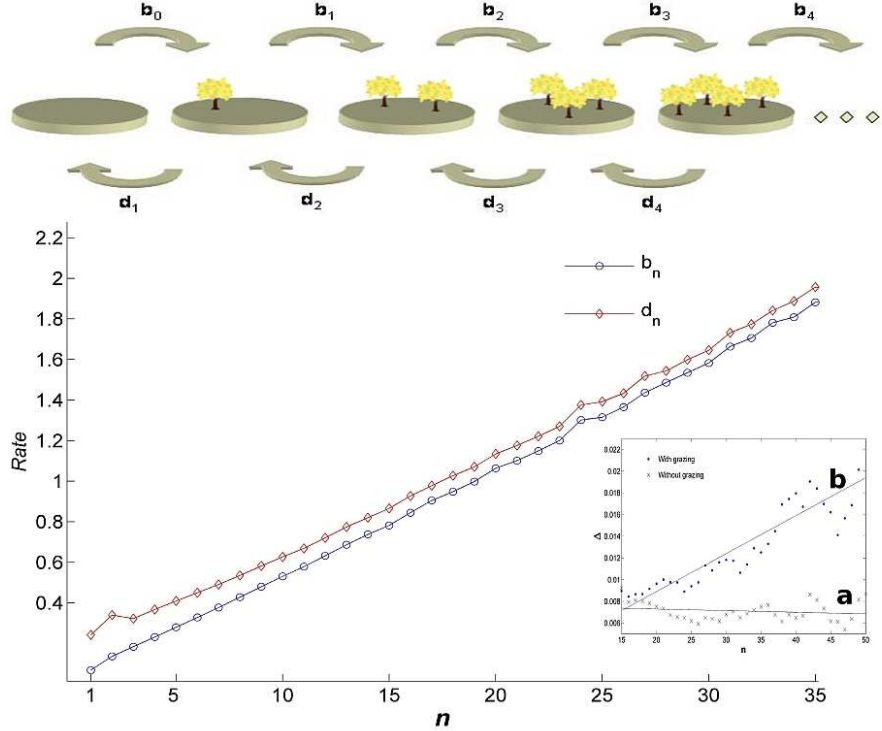}
  \caption{
  Tracing the stochastic simulation along time, the rates in which clusters of size $n$
  grow by one biomass unit, $b_n$, or decay by one unit, $d_n$,
  were monitored (see methods, here the parameters are $f^* = 0.1, \ r =3$). Neglecting merging and splitting of colonies,
  where size is changed by more than one unit, one gets the
  Markov birth-death process illustrated in the upper panel. In
  the lower panel  the birth-death rates  (in arbitrary units) are shown
  in the absence of grazing pressure; the rate of both processes grows linearly with the
  colony's size, where the death rate is always larger by a constant $\Delta$.  In the inset, the
difference $\Delta(n) \equiv d_n-b_n$ is presented without  (a)  and
with (b) grazing pressure together with a linear fit. While
$\Delta(n)$ is approximately  constant without grazing, it shows, on
average, linear growth where grazing is present. This gives the
ground for the approximation $\Delta(n) =\alpha n + \Delta_0$, where
$\alpha$ increases with grazing pressure.  }\label{fig1}

\end{figure*}

Figure 1 shows the dependance of $b_n$ and $d_n$ on $n$. Apparently,
both $b_n$ and $d_n$ grow linearly with $n$. However, the death rate
is larger by  a constant $\Delta$, i.e.,
\begin{equation}
  b_n = \gamma n
 \qquad  \qquad \qquad  d_n = \gamma n + \Delta.
\end{equation}
Using the local balance equation  the steady state may be obtained
analytically:
\begin{equation}\label{solution}
    \pi_n = \frac {b_0 \pi_0} {d_n} \prod_{m=1}^{n-1} \frac {b_m}{d_m} = \frac {b_0 \pi_0} {\gamma n + \Delta} \prod_{m=1}^{n-1} \frac {\gamma m}{\gamma m +
    \Delta},
\end{equation}
where $\pi_0$ is determined by the  normalization condition. The
product  (\ref{solution}) may be calculated exactly, and then
approximated for $n>1+\frac {\Delta}{\gamma}$ to yield,
\begin{equation}\label{solution1}
    \pi_n = \frac {b_0 \pi_0}{\gamma}\frac{\Gamma  ( n ) \Gamma  ( 1+{\frac {\Delta}{\gamma}} )} {  \Gamma( n+1+{\frac {\Delta}{\gamma}} )} \approx A n^{-(1+\frac{\Delta}{\gamma})},
\end{equation}
where $A$ is a normalization constant. Comparison between the
distributions obtained from the cellular automata simulations and
the power-law (\ref{solution1}) using the calculated values of
$\gamma$ and $\Delta$  shows excellent agreement (Figure 2, upper
panel).   This supports the approximation made by replacing the
general Markov chain with a birth-death process.

The monotonic increase of event rates with cluster size emerges,
thus, as the dynamical law that generates criticality. This, in
fact, is the  law of proportionate effect  \cite{ziff}, introduced
many years ago in order to explain the Pareto distributions in
economic systems \cite{gibrat, simon, stanley}. Similar mechanisms
were  shown to underly  surname statistics  \cite{zm1}, relative
species abundance \cite{hub}, urban population size  \cite{sorin}
and many other observations. Although here the geometric random walk
is biased towards extinction, the multiplicative nature of the
process guarantees power-law distributions, where the slope is
determined by the bias $\Delta$. The results presented here are for
facilitation radius $r=3$; testing the model under different
conditions,  the slope $\beta$   was found to increase with  $r$.
Approximating the real Markov chain  as a birth-death process turns
out to be exceptionally good;  for all cases tested by us ($r=2,
3,4$) the slope obtained from $\Delta$ and $\gamma$ differs only by
$5-10  \% $ from the slope measured directly on the  cellular
automata.

As indicated by the field data analyzed by  K\'{e}fi et. al., patch
distribution deviates from power-law in the  presence of grazing
pressure. A truncated power-law has been suggested in that case,
where the strength of the  grazing  determines the typical patch
size above which the distribution crosses over to exponential decay.
Introducing grazing into our model through a density independent
mortality rate (see methods), the same truncated power-law has been
obtained. Mapping to a birth-death Markov chain, the only
qualitative difference is that, under grazing, the leftward bias
$\Delta$ grows linearly with the patch size, i.e.,  $\Delta =
\Delta_0 + \alpha n$ (See inset of Figure 1). Substituting  in
(\ref{solution}) one attains,
\begin{equation}\label{solution2}
    \pi_n = \frac {b_0 \gamma \pi_0}{(\gamma+ \alpha)^2}\frac{\Gamma  ( n ) \Gamma  ( 1+{\frac {\Delta_0}{\gamma + \alpha}} )} {  \Gamma( 1+n+{\frac {\Delta_0}{\gamma+\alpha}} )}
\left( \frac{\gamma}{\alpha+\gamma} \right)^{n}.
\end{equation}
In the limit of $n \gg 1$ this expression becomes the truncated
power-law,
\begin{equation}\label{solution3}
    \pi_n = An^{-\left(1+\frac{\Delta_0}{\alpha+\gamma}\right)} e^{-n/n_0},
\end{equation}
where $n_0 \equiv \left[ ln(\frac{\alpha+\gamma}{\gamma})
\right]^{-1}$ is the patch size above which $\pi_n$ decreases faster
than a power law. The equilibrium distributions (\ref{solution1})
and (\ref{solution3}) fit perfectly the power law and truncated
power law obtained from  direct numerics, as seen in fig. 2.

\begin{figure*}
 \includegraphics[width=13cm]{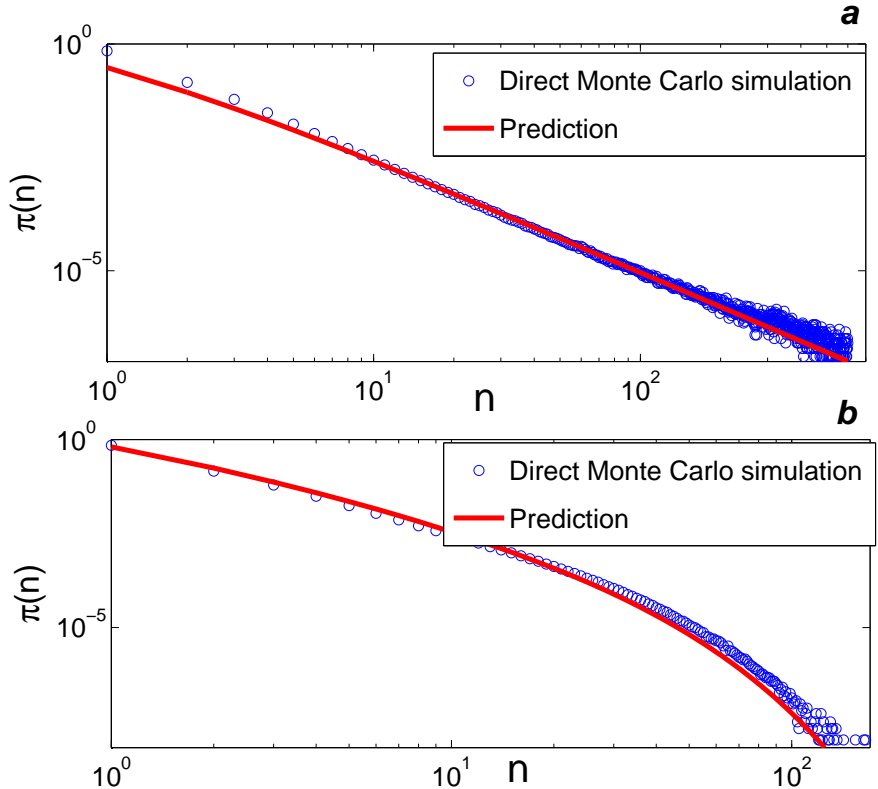}
  \caption{ The equilibrium
probability $\pi_n$ is presented vs. $n$ on a log-log scale, showing
power-law behavior in the absence of grazing [upper panel, (a)]. The
circles are results attained from the stochastic cellular automata,
where the full line corresponds to the analytic expression
(\ref{solution1}) for the solution of the birth-death process. The
lower panel (b) shows the same in the presence of grazing  ($G=0.6$,
see methods), where the  truncated power law attained from the
simulations is compared with (\ref{solution3}). The relevant
parameters $\alpha \ \gamma$ and $\Delta_0$ are those extracted by
linear regression from Figure 1.}\label{fig2}
\end{figure*}

The main characteristic that distinguishes endogenous from exogenous
cluster formation mechanisms is the dynamic of the system. If
spatial heterogeneity plays a major role and determines the location
and the size of large colonies, the chance of such a colony to
disappear due to demographic stochasticity is exponentially small
\cite{mac}, so in practice, on experimental time scales,  large
colonies never disappear. This feature was used  by Vandermeer et.
al. \cite{vandermeer}, who utilized the appearance/disappearance of
large ant colonies as a manifestation of endogenous population
dynamics. Our mapping of the system into a Markov chain allows us to
quantify these characteristics and to establish a connection between
the steady-state distribution and the dynamics of a single cluster.

We define $\tau_n^m$ as the mean first passage time \cite{book_fp}
from some initial cluster size $n$ to some smaller size $m$. Since
the mean time for a cluster of size $n$ to undergo a birth/death
step is $1/(b_n+d_n)$, the mean first passage times should satisfy:
\begin{equation}\label{tau_equation1}
    \tau_{n}^{m}=\frac{1}{b_n+d_n}+\frac{b_n}{b_n+d_n}\tau_{n+1}^{m}+\frac{d_n}{b_n+d_n}\tau_{n-1}^{m}.
\end{equation}
Substituting $b_n=\gamma n$ and $d_n=\gamma n + \alpha_n +
\Delta_0$, we get, for $n \gg 1$:
\begin{equation}\label{tau_equation1}
    \tau_n^m=\frac{1}{\alpha} \ln \left( \frac{\alpha (n-m) +
    \Delta_0}{\Delta_0}
    \right).
\end{equation}
In that limit, the first passage time depends only on the difference
$n-m$.  In the absence of grazing ($\alpha\rightarrow0$), we get
$\tau_{n-m}= (n-m)/\Delta_0$.

\begin{figure*}
\includegraphics[width=13cm]{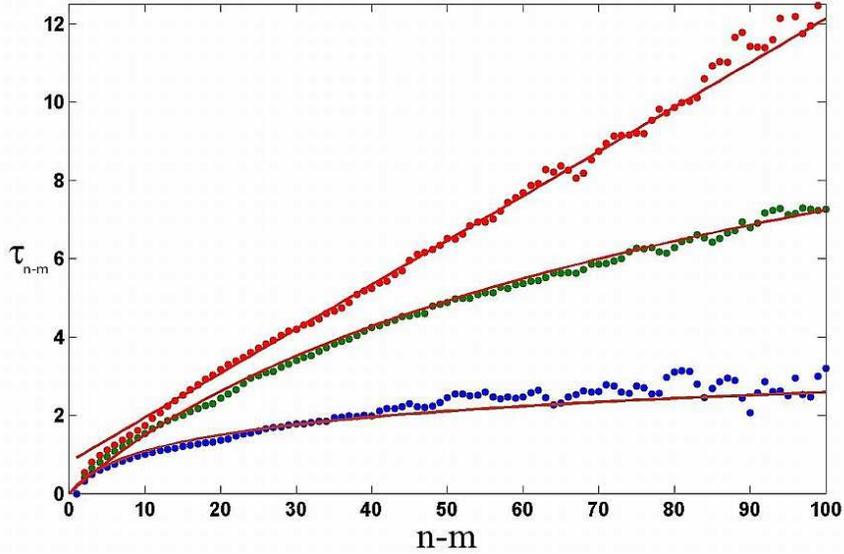}
\caption{ $\tau_{n-m}$, The average time elapsed until a colony of
size $n$ reaches a smaller size $m$ at the first time is plotted
against $n-m$ for different environmental conditions. In the absence
of grazing ($G=0$, red circles) the stability of a colony grows
linearly with its size. Under  grazing pressure ($G= 0.4$, green and
$G=0.6$, blue) the time to extinction grows only logarithmically
with the patch size. Datasets were taken directly from the
simulation of the cellular automata with $r=3$ and $f^* = 0.1$, and
are fitted (full lines) to the predictions (\ref{tau_equation1})
using nonlinear curve fitting. }
\end{figure*}

Evidentally, under grazing pressure, the dependence of $\tau_{n-m}$
on $n-m$ is logarithmic, while in the absence of grazing, the
dependence is linear, as demonstrated in Figure 3. Tracking the
temporal variations of colonies of different sizes (e.g., using
satellite images taken yearly or seasonally) is thus an effective
tool for detecting desertification. If large patches remain large,
and their chance to reach a fixed, small size within some time frame
decays exponentially with their size, it is plausible that the
spatial structure reflects spatial heterogeneity, effects of local
topography and so on. If the rate of disappearance for large
clusters depends linearly on size, it indicates that positive
feedback plays a major role in the clustering process. Grazing, and
other processes that lead to desertification, are characterized by
logarithmic dependence on size differences. These three
characteristic dynamics - exponential, linear, and logarithmic -
differ strongly from each other, and thus temporal tracking provides
a very reliable indication of the state of the system.

It is interesting to note that, as firm size is governed by the same
dynamical law, similar techniques are used for financial risk
assessment. A Firm  defaults when the value of its liabilities
exceeds its assets. Merton \cite{merton} estimated the distance to
default (DD) by modeling a firm assets as  a geometric random walk,
where the "step size" is proportional to the  volatility. The
expected default frequency \cite{hh} (EDF), which is the critical
parameter that determines credit risk,  is in fact an estimation of
the typical time window needed for the geometric  random walker to
reach the default point; this quantity is just the inverse of  the
first passage time considered here.

The models of  \cite{scanlon,kefi,vandermeer} all use a fixed
fraction of occupied cells $f^*$, i.e.,  a fixed total population.
The  differences in colony statistics quantify, in fact, the
relative importance of the density-dependent growth factors (local
facilitation, Allee effect, schooling) versus density-independent
resources, like annual precipitation or the availability of food. If
the patch distribution is Pareto-like with a relatively small slope
$\beta$, the system admits larger patches; this implies that the
effect of positive feedback is strong enough to oppose stochasticity
(purely stochastic dynamics lead to Poisson-like statistics). A
truncated power-law, on the other hand, shows that
density-independent factors control the overall carrying capacity.
Given two systems with the same tree coverage  $f^*$ but with
different cluster statistics, the scale-free one is more robust
against environmental changes (e.g., decreased annual rainfall) than
the system characterized by a truncated power law. On the other
hand, a strong Allee effect may result in irreversible transitions
if the density falls below some critical level. The dynamical law
presented here allows one to identify the transition between these
states when the total population is still fixed; qualitative changes
in the stability of large patches - from linear to logarithmic - may
thus serve as  an important precursor of catastrophic shifts
\cite{cat} like desertification in arid ecosystems.

\begin{acknowledgments}
 This work was supported by the EU 6th framework CO3 pathfinder.
 A.M. acknowledge the financial support of the Israeli  Center for
Complexity Science.
\end{acknowledgments}

\end{document}